\documentclass[11pt,a4paper,fleqn]{article}
\usepackage[DIV12]{typearea}
\usepackage{amsmath,amsfonts,amssymb}
\usepackage{graphicx}
\usepackage{cite}
\usepackage{mathrsfs}
\usepackage{bbm}
\usepackage{pifont}
\usepackage{units}
\usepackage{longtable}
\usepackage[small,loose]{subfigure}  
\usepackage{pst-all}
\usepackage{cancel}
\usepackage{longtable}
\usepackage{cite} 

\DeclareMathOperator{\re}{Re}

\newcommand{\CenterObject}[1]{\ensuremath{\vcenter{\hbox{#1}}}}
\newcommand{\CenterEps}[2][1]{\ensuremath{\vcenter{\hbox{\includegraphics[scale=#1]{#2.eps}}}}} 

\newcommand{\E}[1]{\ensuremath{\mathrm{E}_{#1}}} 
\newcommand{\G}[1]{\ensuremath{\mathrm{G}_{#1}}}
\newcommand{\SO}[1]{\ensuremath{\mathrm{SO}(#1)}}
\newcommand{\SU}[1]{\ensuremath{\mathrm{SU}(#1)}}
\newcommand{\U}[1]{\ensuremath{\mathrm{U}(#1)}}
\newcommand{\Z}[1]{\ensuremath{\mathbbm{Z}_{#1}}} 

\title{Stringy surprises}

\author{Michael \textsc{Ratz}%
}

\advance \headheight by 3.0truept       

\begin{document}

\thispagestyle{empty}
\begin{titlepage}

\begin{flushright}
TUM-HEP 753/10
\end{flushright}

\vspace*{1.0cm}

\def\oldfootnote{\thefootnote}
\renewcommand\thefootnote{\fnsymbol{footnote}}
\begin{center}
\Huge\textbf{Stringy surprises}\footnote{Invited proceedings prepared for the
Yukawa International Seminar Symposium in Kyoto 2009. Based also on invited
plenary talks at the String Pheno 08 (Philadelphia), SUSY 08 (Seoul), GUTs and
strings 09 (Hamburg), GDR Terascale meeting 09 (Grenoble), Planck 09 (Padova)
and String Pheno 09 (Warsaw) conferences as well as in a parallel session talk
given at ICHEP 08 (Philadelphia) and a talk at the Aspen Center for Physics in
2009.}
\end{center}
\renewcommand\thefootnote{\oldfootnote}
\setcounter{footnote}{0}
\vspace{1cm}
 \center{
\textbf{
Michael Ratz
}
}
\\[5mm]
\begin{center}
\textit{\small
Physik-Department T30, Technische Universit\"at M\"unchen, \\
James-Franck-Stra\ss e, 85748 Garching, Germany
}
\end{center}

\date{}
\vspace{1cm}

\begin{abstract}
There are many conceivable possibilities of embedding the MSSM in string theory.
These proceedings describe an approach which is based on grand unification in
higher dimensions. This allows one to obtain global string-derived models with
the exact MSSM spectrum and built-in gauge coupling unification. It turns out
that these models exhibit various appealing features  such as (i) see-saw
suppressed neutrino masses,  (ii) an order one top Yukawa coupling and
potentially realistic flavor structures,  (iii) non-Abelian discrete flavor
symmetries relaxing the supersymmetric flavor problem, (iv) a hidden sector
whose scale of strong dynamics is consistent with TeV-scale soft masses, and (v)
a solution to the $\mu$-problem.  The crucial and unexpected property of these
features is that they are not put in by hand nor explicitly searched for but
happen to occur automatically, and might thus be viewed as  ``stringy
surprises''.
\end{abstract}

\end{titlepage}


\section{Goals of string model building}
\label{sec:WhyStrings}

The standard model (SM) of elementary particle physics is remarkably successful
in describing experiments. There are three main reasons for going beyond the SM:
\begin{dingautolist}{'300}
 \item observational: \label{observational}
 neither the observed cold dark matter nor the baryon
 asymmetry can be explained in the SM;
 \item conceptual: \label{conceptual}
 the SM is based on quantum field theory, in which, however,
 it appears difficult to incorporate gravity;
 \item aesthetical: \label{aesthetical}
 the structure and the parameters of the SM ask for a simple,
 arguably more fundamental explanation.
\end{dingautolist}
Solid observations contradicting the SM so far are mostly astrophysical and/or
cosmological. There are many ways to extend the SM such as to explain these
observations; perhaps even too many. One might therefore argue that one should
search for theoretical guidelines that, in a way, reduce the arbitrariness in
model building. In these proceedings, the guideline will be the requirement that the
extension of the SM should be embedded into string theory, which is believed to
unify quantum gauge theory with gravitation. This choice is motivated by the
above reasons \ref{conceptual} and \ref{aesthetical}, and sort of ignores the
most concrete arguments \ref{observational} for going beyond the SM. This
approach builds on the observation that the gauge couplings appear to meet in
the minimal supersymmetric extension of the SM, the MSSM, at the scale 
\begin{equation}
 M_\mathrm{GUT}~=~\text{few}\cdot 10^{16}\,\mathrm{GeV}\;,
\end{equation}
and that the four-dimensional Planck scale $M_\mathrm{P}$ is numerically not too
far from $M_\mathrm{GUT}$. Explanations of the smallness of the neutrino masses
often rely on a similarly high scale. Even more, the fact that one generation of SM matter
fits into the $\boldsymbol{16}$-plet of \SO{10} is interpreted as strong
evidence for unification along the exceptional chain~\cite{Olive:1981tb}{}
\begin{equation}
 \E3=G_\mathrm{SM}~\subset~\E4=\SU5~\subset~\E5=\SO{10}
 ~\subset~\E6~\subset~\E7~\subset~\E8\;,
\end{equation} 
which is beautifully realized in the heterotic string
~\cite{Gross:1984dd,Gross:1985fr}{}~(cf.\ the discussion
in~\cite{Nilles:2004ej}{}). Here $G_\mathrm{SM}$ denotes the SM gauge
group,
\begin{equation}
 G_\mathrm{SM}~=~\SU3_C\times\SU2_\mathrm{L}\times\U1_Y\;,
\end{equation}
The main purpose of these proceedings is to show that the emerging route from the SM
to string theory, the ``heterotic road'', has particularly promising features. 

One of the main motivations of building a string model is as follows.
A string-derived model has to be `complete' in the following sense: once one has
obtained a globally consistent construction that reproduces the SM in its
low-energy limit, unlike in field theory one cannot `amend' it by extra
ingredients such as extra hidden sectors, further states or additional
interactions. Instead, we have to live with what string theory gives us. 
In particular, solutions to the usual open questions, such as the strong
CP problem, have to be already included in a global string-derived model. Since
spectrum and couplings are, in principle, calculable, one might hope to arrive
thus at non-trivial predictions. The strategy would then be to
\begin{dingautolist}{'312}
 \item first construct a model that reproduces the SM in its
  low-energy limit and
 \item then identify solutions to long-standing puzzles in this construction.
\end{dingautolist}
The main problem with this strategy is that the first step is highly
non-trivial. 
In fact, the first item \ding{'312} has been around for a rather long time;
already in 1987 L.~Ib\'a\~{n}ez made the statement more
concise~\cite{Ibanez:1987dw} by defining a sort of ``wish list'':
\begin{enumerate}
\item chirality;
\item gauge group contains (and can be broken to) $\SU3\times\SU2\times\U1$;
\item $N=1$ supersymmetry in $d=4$;
\item contains standard quark-lepton families;
\item contains Weinberg-Salam doublets;
\item three quark-lepton generations;
\item proton is sufficiently stable 
	($\tau_p\gtrsim 10^{3\red{\xcancel{\black0}4}}\,$years);
\item  	correct prediction
	$\sin^2\theta_\mathrm{W}\simeq0.2\red{\xcancel{\black1}}\red{3}$, $M_X\simeq
	M_\mathrm{P}\simeq10^{18}\,\mathrm{GeV}$;
\item no exotic gauge boson with mass 
	~$\lesssim\red{\xcancel{\black100\,\mathrm{GeV}}~1\,\mathrm{TeV}}$
nor fermions 
	$\lesssim\red{\xcancel{\black40}~100}\,\mathrm{GeV}$;
\item no flavour-changing neutral currents;
\item {$\red{\xcancel{\black \text{massless}}}$} (or so) left-handed neutrino;
\item weak CP violation exists;
\item potentially realistic Yukawa couplings (fermion masses);
\item $\SU2\times\U1$ breaking feasible;
\item small supersymmetry breaking;
\item \dots
\end{enumerate}
It is remarkable that the experimental situation did not change much at the
qualitative level since then (the updates to the traditional wish list are
marked in red). Clearly,  if one was to go back from 1987 by additional 22
years, analogous wish lists would have changed dramatically. Yet, despite the
relatively long time of rather little changes to the wish list, string theory
did not yet give us a clear answer.  In fact, so far only few models have been
found which come close to the (MS)SM. Some of the most common problems are that
concrete string compactifications predict unwanted states that cannot be
decoupled, so-called chiral exotics, and/or unrealistic interaction patterns
such as a hierarchically small top Yukawa coupling.

One important comment to make in this context is the following: if a model
predicts wrong quantum numbers, it is certainly ruled out.  On the other hand, a
model is definitely not ruled out if it does not comply with the currently most
popular ideas of moduli fixing.  In other words, it is by far more likely that
string theorists have missed some possibilities for moduli stabilization than
that experimentalists have overlooked some chiral exotics at LEP 2 or the
Tevatron. Therefore, our strategy is to seek for models that give rise to the
right states and interaction patterns, and to approach the really tough
questions like the breakdown of supersymmetry, moduli stabilization and the
vacuum energy in this class of models in a second step. As we shall see later,
this strategy has led to novel ideas in moduli fixing and explaining the
hierarchy between the Planck and electroweak scales.

These proceedings might be viewed as an addendum to the earlier
reviews~\cite{Ratz:2007my,Nilles:2008gq,Raby:2008gh}{}, where heterotic orbifold
compactifications have been described that exhibit the exact MSSM spectra at low
energies.  Before entering the details, a couple of disclaimers
and apologies are in order:
\begin{enumerate}
 \item this is \textbf{not} going to be a complete survey of all
 attempts to find the MSSM;
 \item the focus will be on models with the exact MSSM spectrum at low energies
 and built-in gauge coupling  unification;\footnote{Recently intersecting $D$
 brane models have been constructed which possess the chiral spectrum of the
 MSSM \cite{Gmeiner:2007zz,Gmeiner:2008xq}.  These models will not be discussed
 because there gauge coupling unification appears to be an accident rather than
 built in, and the prejudice in these proceedings is unification.}
 \item only globally consistent string models will be discussed;
 \item the focus is on models with a clear geometric
 interpretation;\footnote{There are also very promising models based on the free
 fermionic construction in the literature (for a review see
 \cite{Cleaver:2007ek}{}). Whether or not these constructions have a geometric
 interpretation is controversial~\cite{Donagi:2008xy}{}.}
 \item there are alternatives, satisfying the above criteria,
 which will also not be discussed in detail
 \cite{Bouchard:2005ag,Braun:2005nv,Kim:2007mt}.
\end{enumerate}

\section{Exact MSSM spectra from heterotic orbifolds}

The focus of these proceedings will be on MSSM models based on the \Z6-II 
orbifold~\cite{Buchmuller:2005jr,Buchmuller:2006ik,Lebedev:2006kn,Lebedev:2006tr,Lebedev:2007hv,Lebedev:2008un}{}.
They were obtained by
marrying the bottom-up idea of orbifold
GUTs~\cite{Kawamura:1999nj,Kawamura:2000ev,Altarelli:2001qj,%
Hall:2001pg,Hebecker:2001wq,Asaka:2001eh,Hall:2001xr,Burdman:2002se}\ (for a
review, see e.g.\ \cite{Quiros:2003gg}) to the orbifold compactifications of the
heterotic
string~\cite{Dixon:1985jw,Dixon:1986jc,Ibanez:1986tp,Ibanez:1987sn,Casas:1987us,Casas:1988hb,Font:1988tp,Font:1988nc}{}.
A key ingredient of these constructions is a non-trivial gauge group
topography~\cite{Forste:2004ie}{}, i.e.\ different gauge groups are realized at
different positions in compact space. More precisely, the bulk gauge group
$\E8\times\E8$ gets broken to different subgroups, which will be referred to as
``local groups'', at different orbifold fixed points or planes. The effective
gauge group after compactification is given by the intersection of the various
local groups in $\E8\times\E8$. By demanding that one \E8 factor gets broken to
$G_\mathrm{SM}$,
one is then lead to the picture of ``local grand unification''
(LGU)~\cite{Buchmuller:2005sh,Buchmuller:2007qf,Ratz:2007my}{}. Here, the local
gauge groups are larger than $G_\mathrm{SM}$,
\begin{equation}
 G_\mathrm{local}~\supset~G_\mathrm{SM}\;;
\end{equation}
hence these groups are precisely those discussed in the context of (4D) grand
unification,
\begin{equation}\label{eq:GlocalGUT}
 G_\mathrm{local}~=~
 \SU5,~G_\mathrm{PS},~\SO{10}
 \quad\text{etc.}\;,
\end{equation}
where $G_\mathrm{PS}=\SU4\times\SU2_\mathrm{L}\times\SU2_\mathrm{R}$ is the
Pati-Salam group~\cite{Pati:1974yy}{}.
The key ingredient of the LGU scheme is that states
confined to a region with a GUT symmetry, equation~\eqref{eq:GlocalGUT},
necessarily furnish complete representations of that symmetry. On the other
hand, bulk fields turn out to "feel" symmetry breaking everywhere, and hence
come in split multiplets.

Although LGU scenarios can be obtained in the context of field theory, we will
argue that it is advantageous to embed the LGU scheme into string theory. Apart
from the reasons described in section~\ref{sec:WhyStrings}, strings are --
unlike gauge field theories in more than four dimensions -- well behaved, i.e.\
they are believed to be UV complete. On the practical side, a stringy
computation of the spectrum of a given orbifold model is straightforward whereas
in field theory it is very hard to figure out what the states at the fixed
points are. Moreover, in string-derived models, all anomalies, including those
in higher dimensions, cancel. This has been verified explicitly in an
example~\cite{Schmidt:2009bd}{}; looking a the complexity of the constraints it
appears very hard to construct a model in the bottom-up approach where they all 
are satisfied.

\begin{figure}[h]
\centerline{\CenterEps{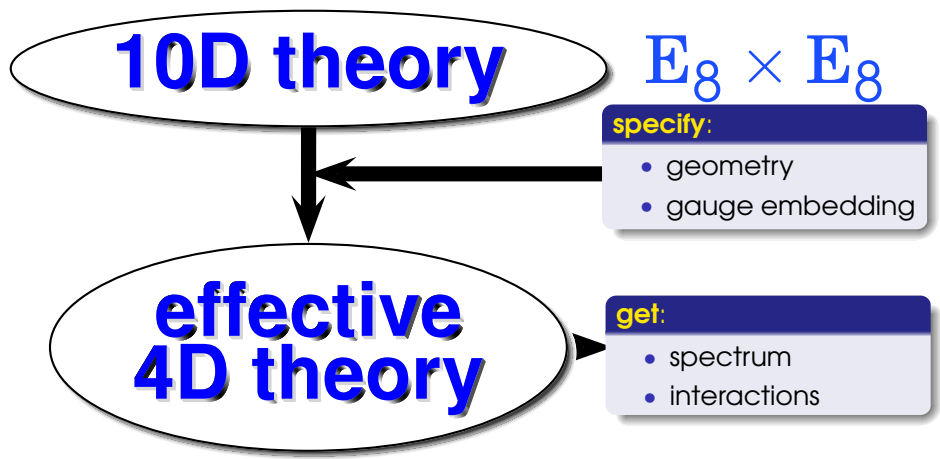}}
\caption{Orbifold model building.}
\label{fig:stringmodelbuilding}
\end{figure}

Let us briefly outline how such stringy orbifold compactifications work. (For a
detailed description and recipes on orbifold computations
see~\cite{Vaudrevange:2008sm,RamosSanchez:2008tn}{}; for the \Z6-II case
see~\cite{Buchmuller:2006ik}{}.) A model is defined by the
geometry and the so-called gauge embedding
(figure~\ref{fig:stringmodelbuilding}). Notice that a model has many vacua with
very different phenomenological properties. The geometry of a string $\Z{N}$
orbifold is defined by a 6D torus $\mathbbm{T}^6$ and a symmetry operation
$\theta$ which can be modded out. This symmetry operation is to be embedded into
the gauge degrees of freedom. This is described by the so-called gauge shift
$V$. Moreover, the torus translations $e_\alpha$ can be
associated to discrete Wilson lines $W_\alpha$, which have to comply with the
discrete symmetry operation $\theta$.
Consistency conditions then limit the possible number of models to a finite
number. A complete classification of all gauge embeddings has first been
attempted for the \Z3 orbifold~\cite{Giedt:2000bi,Giedt:2003an}{}. The main
problem is that there is a huge redundancy in the shift $V$ and Wilson lines. In
orbifold models, two sets $(V,W_\alpha)$ and $(V',W_\alpha')$ are equivalent if they
are related by Weyl reflections. If they differ by vectors in the root lattice of $\E8\times\E8$,
$\Lambda_{\E8\times\E8}$, they are equivalent, or fall into a small number of
equivalence classes, called brother models in~\cite{Ploger:2007iq}{}. However, the Weyl
group of \E8 is enormously large, the number of elements is 696729600, so that
in practice it is impossible to check whether two shifts are equivalent or not.
Giedt's method \cite{Giedt:2000bi,Giedt:2003an} allows one to eliminate these
redundancies to a large extent, but not completely. To obtain the true number of 
models, in~\cite{Lebedev:2008un}{} a statistical method, based on proposals made
in a different context~\cite{Dienes:2006ut,Dienes:2006ca}{}, has been described.
There shifts and Wilson lines are generated randomly, and the spectra are
computed. One builds up sets of models by the following procedure: generate a
first model. Then generate randomly further models and add them to the set as
long as they are not already contained in this set; if the model was already
present, terminate. The criterion if two models are equivalent or not is taken
to be whether or not the spectra coincide; this underestimates the true number
of models somewhat. The size of thus generated model sets goes as $\sqrt{N}$,
where $N$ denotes the number of inequivalent models. Using this strategy, one
finds that the \Z6-II orbifold admits roughly $10^7$ inequivalent gauge
embeddings.

Let us now come to how the promising models with the exact MSSM spectra were
found. The search strategy was based on the concept of LGU, as explained above.
In the context of the heterotic orbifolds, this means that one should look at
compactifications that exhibit fixed points with local GUT groups and localized
GUT representations, which eventually give rise to complete SM representations.
The simplest way to obtain a three-generation model is to look at models in
which there are three fixed points with an \SO{10} symmetry and localized
$\boldsymbol{16}$-plets \cite{Buchmuller:2004hv}{}. However, it turns out that
stringy consistency conditions (such as modular invariance) are so restrictive
that in all settings of this type one has to buy extra states which imply that
one either has to allow for chiral exotics or play with the normalization of
hypercharge, thus giving up the simple picture of MSSM gauge coupling
unification~\cite{Buchmuller:2006ik,Lebedev:2006kn}{}. 

Since the idea of three sequential families does not work smoothly, one has to
look for alternatives. The perhaps simplest possibility is to go for ``$2+1$
family models'', i.e.\ settings where two families are explained as completely
localized $\boldsymbol{16}$-plets while the third family comes from `somewhere
else'. Models of this type have first been studied in the context of
string-derived Pati-Salam models~\cite{Kobayashi:2004ud,Kobayashi:2004ya}{}. In
what follows, we shall focus on MSSM models with this structure
\cite{Buchmuller:2005jr,Buchmuller:2006ik,Lebedev:2006kn,Buchmuller:2007qf,Lebedev:2007hv,Buchmuller:2008uq}{}.
It turns out that $2+1$ family models are indeed promising:
\begin{enumerate}
 \item one can find $\mathcal{O}(100)$ models with the chiral MSSM spectra,
 which denote the so-called ``heterotic mini-landscape'';
 \item exotics are vector-like w.r.t.\ $G_\mathrm{SM}$ and can be decoupled
 consistently with vanishing $F$- and $D$-terms;
 \item these settings exhibit various additional good features automatically,
 i.e.\ one does not have to search for these features, they are simply there.
\end{enumerate}
It is the last point which motivates the title of these proceedings, and which
will be the focus of the subsequent discussion.

Before discussing the surprising features, let us note that, in order to get rid
of the vector-like exotics, one has to switch on VEVs of certain SM singlet
fields $s_i$. Giving VEVs to states localized at certain fixed points
corresponds to resolving or `blowing up' the respective singularity (for a
recent discussion see~\cite{GrootNibbelink:2007pn,Nibbelink:2008tv}{}). Often
one can blow up an orbifold completely, thus arriving at a smooth Calabi-Yau
space. However, in  the models we shall discuss it turns out that a complete
blow-up always destroys some phenomenologically important features of the
models, for instance breaks hypercharge at a high
scale~\cite{Nibbelink:2009sp}{}. This fact has been interpreted in different
ways. The authors of \cite{Nibbelink:2009sp}{} regard it as fine tuning if not
all singularities are blown up. On the other hand, string theory is known to be
well-behaved at the orbifold point since the very first papers on this
subject~\cite{Dixon:1986jc}{}. Even more, the orbifold point denotes a
symmetry-enhanced configuration in moduli space, and it is well known that
moduli tend to settle at such points~\cite{Dine:1998up,Kofman:2004yc}{} (for a
recent field-theoretic example demonstrating this
see~\cite{Buchmuller:2009er}{}). This is because these are typically stationary
points of the scalar potential. It is also clear that in the presence of a 4D
Fayet-Iliopoulos (FI) term, not all fields can reside at the orbifold point.
Instead, one has to go to a `nearby vacuum' in which the FI term is
canceled~\cite{Atick:1987gy}{}. In such a situation, some fields get driven
somewhat away from the orbifold point while other stay there. Of course, these
arguments do not tell us why only some SM singlets attain VEVs, yet they might
nevertheless allow us to give a preference for so-called partial against full
blow-ups.

The search for MSSM models in the \Z6-II orbifold has been completed
in~\cite{Lebedev:2008un}{}. It turns out that there are $\mathcal{O}(100)$
models without the $2+1$ family structure, still giving rise to the exact MSSM
spectra.  In this scan, about $5\cdot10^6$ out of a total of $\sim10^7$
inequivalent models has been analyzed.  Most MSSM candidates are based on two
and some on one local GUTs (see table~\ref{tab:2or3WL}). Interestingly, although
the subset of models with 2 equivalent families, i.e.\ the models with two out
of three possible non-trivial Wilson lines, is very small, only about $3\cdot
10^4$ out of $10^7$ models have this structure, the majority of MSSM candidates
is based on 2 Wilson lines.
\begin{table}[h]
\begin{center}
\begin{tabular}{c|c|r|r}
 local GUT & ``family'' &  2~WL & 3~WL \\
\hline
 &&&\\[-0.4cm]
 $\E6$          & $\boldsymbol{27}$ & $14$  & $53$ \\[0mm]
 \SO{10}      & $\boldsymbol{16}$ & $87$  &  $7$ \\[0mm]
\SU6         & $\boldsymbol{15}$+$\boldsymbol{\bar 6}$ &  $2$  &  $4$ \\[0mm]
\SU5         & $\boldsymbol{10}$ & $51$  & $10$ \\[0mm]
rest &                 & $39$  &  $0$ \\
 \hline
 total &                 & $193$ & $74$ \\
 \hline
\end{tabular}
\end{center}
\caption{Result of a random scan for MSSM models in the \Z6-II orbifold. ``2~WL''
and ``3~WL'' means that two or three Wilson lines are switched on.}
\label{tab:2or3WL}
\end{table}
Only a small subset of 39 candidates does not exhibit local GUT structures at
all. This might be interpreted as evidence for the importance of incorporating
elements of grand unification into string model building. 

\section{Phenomenological properties}

Let us now discuss some of the most important phenomenological properties oof
these models. We will mainly focus on the 2WL models, since they are, as of now,
better explored.
They fall into two classes, depending on the shift; it can
be either
\begin{subequations}
\begin{equation}
 V_\mathrm{KRZ}
 ~ = ~ 
 \left(\frac{1}{3},\frac{1}{3},\frac{1}{3},0,0,0,0,0\right)
 \left(\frac{1}{6},\frac{1}{6},0,0,0,0,0,0\right)
\end{equation}
or  
\begin{equation}\label{eq:VBHLR}
 V_\mathrm{BHLR}
 ~=~ 
 \left(\frac{1}{2},\frac{1}{2},\frac{1}{3},0,0,0,0,0\right)
 \left(\frac{1}{3},0,0,0,0,0,0,0\right)\;.
\end{equation}
\end{subequations}
$V_\mathrm{KRZ}$ has been first used in the context of  Pati-Salam
models~\cite{Kobayashi:2004ud,Kobayashi:2004ya}{} while the first MSSM models in
the \Z6-II orbifold were based on 
$V_\mathrm{BHLR}$ \cite{Buchmuller:2005jr,Buchmuller:2006ik}{}.

\subsection{Neutrino masses}

\begin{figure}[h]
\centerline{\CenterEps{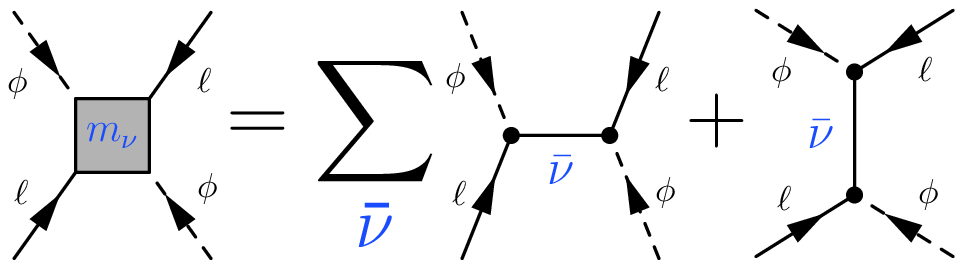}}
\caption{See-saw.}
\label{fig:see-saw}
\end{figure}

One of the most striking observations supporting the picture of the great desert
between the electroweak and GUT scales comes from neutrino masses, which are
known to be small, 
\begin{equation}
 m_{\nu}~\lesssim~0.1\,\mathrm{eV}\;.
\end{equation}
The smallness of $m_\nu$ can, in a very compelling way, be related to the
hierarchy between the GUT and electroweak scales.
The most prominent realization is the see-saw~\cite{Minkowski:1977sc}{},
where the neutrino masses are given by the famous formula
\begin{equation}\label{eq:see-saw}
 m_{\nu}~\sim~\frac{v^2}{M_{\bar\nu}}
\end{equation}
with $v$ and $M_{\bar\nu}$ denoting the electroweak VEV and the mass of right-handed
neutrinos $\bar\nu_i$, respectively. Data, in particular from the atmospheric
neutrino oscillations, seem to indicate that $M_{\bar\nu}$ has to be somewhat below the
GUT scale. It turns out that the mini-landscape has a built-in mechanism to
lower the see-saw scale against the mass scale of vector-like exotics, which can
be argued to be of the order of the GUT or compactification scale. The mechanism
relies on the presence of $\mathcal{O}(100)$ instead of three right-handed
neutrinos $\bar\nu_i$. To understand this, let us explain what a neutrino in
these constructions is. To be specific, we focus on vacua with matter or $R$
parity, some of which have been
explored in~\cite{Lebedev:2006kn,Lebedev:2007hv,Buchmuller:2008uq}. A neutrino is
then simply an $R$-parity odd $G_\mathrm{SM}$ singlet.  To obtain the see-saw
formula~\eqref{eq:see-saw}, one has to integrate out the right-handed
neutrinos. In other words, the effective neutrino masses get contributions from
all neutrinos (figure~\ref{fig:see-saw}).
In the presence of $N_{\bar\nu}>3$ neutrinos, $m_\nu$ gets enhanced against what
one gets in the the 3-neutrino case~\cite{Buchmuller:2007zd}{}, with the
enhancement factor going roughly as $\sqrt{N_{\bar\nu}}$~\cite{Ellis:2007wz}{}.
Because of the contributions of many neutrinos outside the \SO{10}
$\boldsymbol{16}$-plet, the flavor structure of $m_\nu$ is not directly related
to the flavor structure of the quarks and charged leptons. To first
approximation, one gets some flavor anarchy~\cite{Hall:1999sn}{}; deriving
reliable textures in specific vacua along the lines of~\cite{Smirnov:1993af}{}
appears also feasible.

\subsection{Flavor structure}

Let us now take a closer look at the Yukawa couplings of charged fermions. Here,
we focus on the $\mathcal{O}(100)$ models based on $V_\mathrm{BHLR}$
(equation~\eqref{eq:VBHLR}). They turn
out to have the following family structure (up to vector-like states):
\begin{itemize}
 \item $1^\mathrm{st}$ and $2^\mathrm{nd}$ families come from 
 $\boldsymbol{16}$-plets localized at \SO{10} fixed points;
 \item $3^\mathrm{rd}$ family $\bar d$ and $\ell$ (i.e.\ the $3^\mathrm{rd}$
 family $\overline{\boldsymbol{5}}$ in \SU5 language) come from the $T_{2/4}$
 sectors and therefore correspond to states localized on two-dimensional
 submanifolds in compact 6D space;
 \item $3^\mathrm{rd}$ family $\bar u$, $\bar e$ and $q$ as well as the Higgs fields $h_u$
 and $h_d$ are bulk fields, i.e.\ free to propagate everywhere in compact space.
\end{itemize}
Let us discuss implications of these facts at a naive, field-theoretic level.
Yukawa couplings connecting the Higgs fields to matter may be written as overlap
integrals, one could then expect that the couplings of the first two generations
are suppressed by the total 6D volume while the $\tau$ and $b$ Yukawas, $y_\tau$
and $y_b$, are suppressed by the 4D volume transverse to the
two-dimensional submanifold and the top Yukawa $y_t$ is unsuppressed, thus
leading to the hierarchy
\[
 \text{Yukawa couplings of the first two generations}
 ~\ll~
 y_\tau,y_b
 ~\ll~ y_t\;.
\]
Needless to say that this is not against data. It is somewhat surprising that,
at least in the search based on local GUTs, the heterotic string did not allow
us to get MSSM models with three sequential families, where the flavor structure
would have been unrealistic. On the contrary, it forced us to go to models where
the appearance of the third family is somewhat miraculous, but the flavor
structure is qualitatively realistic.

The top Yukawa coupling $y_t$ plays a special role as it is directly related to
the (unified) gauge coupling. At tree level, one obtains an equality between
$y_t$ and the unified gauge coupling~\cite{Buchmuller:2007qf}{}
\begin{equation}
 y_t~=~g\;.
\end{equation}
This relation is subject to various corrections. Apart from the usual 4D
renormalization group running, the most important modifications of this relation
stem from non-trivial localization effects. To discuss these, consider an
orbifold GUT limit in which the \SO4 plane gets large. Here, the right-handed
top quark and the third generation quark doublet are contained in a
hypermultiplet~\cite{Buchmuller:2007qf}{}. The two different $N=1$ components of
this hypermultiplet attain different non-trivial profiles due to the presence of
localized Fayet-Iliopoulos (FI) terms~\cite{Lee:2003mc}{}. As a consequence, the
prediction for $y_t$ at the compactification scale gets reduced against the
gauge coupling $g$, where the suppression depends the geometry of internal
space~\cite{Hosteins:2009xk}{}. On the other hand, the value of $y_t$ at the
compactification or GUT scale translates into a prediction for the ratio of
Higgs VEVs $\tan\beta$. It turns out that the reduction is phenomenologically
welcome, and allows us to obtain moderately large (or even large) $\tan\beta$,
which seem to be favored by phenomenology, in particular by the LEP bound on the
lightest Higgs mass. A rough estimate of the reduction seems to indicate that
rather anisotropic geometries, allowing for an orbifold GUT interpretation, are
favored~\cite{Hosteins:2009xk}{}.

Such anisotropic geometries allow us, at the same time, to reconcile the GUT
scale with the string scale~\cite[footnote~3]{Witten:1996mz}.
This can be accomplished by associating $M_\mathrm{GUT}$ to the inverse of the
largest radius, while all (or most of) the other radii are much smaller. In this
case, the volume of compact space can be small enough to ensure that the
perturbative description of the setting is still appropriate. This idea has been
studied in some detail more recently \cite{Hebecker:2004ce}. The outcome of the
analysis is that the above puzzle can be resolved if the largest radius is by a
factor 50 or so larger than the other radii.
Amazingly, the estimate of the suppression of the top Yukawa coupling reveals
that, in order to obtain phenomenologically attractive values for $\tan\beta$,
one has to go to a rather anisotropic orbifold. This gives further support for this
idea of reconciling the GUT and string scales.\footnote{Power-like running
between the different compactification scales has been analyzed recently
in~\cite{Dundee:2008ts,Dundee:2008gr}{}. In the context of \Z{12} it was found that stringy
threshold corrections and power-like running might be different in orbifolds
with Wilson lines~\cite{Kim:2007jg}{}. These issues deserve to be studied in
more detail~(cf.~\cite{Klaput:2010dg}).}

Another important issue is the flavor structure of the soft supersymmetry
breaking terms. The fact that the two light generations reside at two equivalent
fixed points has important implications. As a consequence, the two light
generations transform as a doublet under a $D_4$ discrete flavor
symmetry~\cite{Kobayashi:2004ud,Kobayashi:2006wq}{}. Therefore, the structure 
of the soft masses is~\cite{Ko:2007dz}{}
\begin{equation}
 \widetilde{m}^2~=~\left(\begin{array}{ccc}
  a &  0 & 0\\
  0 &  a & 0\\ 
  0 &  0 & b
 \end{array}\right)+\text{terms proportional to $D_4$ breaking VEVs}\;.
\end{equation}
This structure is very much reminiscent of the scheme of ``minimal flavor
violation'' (MFV) \cite{Chivukula:1987py,Buras:2000dm,D'Ambrosio:2002ex}, in
which the soft masses are of the form
\begin{equation}\label{eq:MFVansatz}
 \widetilde{m}^2_\mathrm{MFV}~=~a\,\mathbbm{1}+b\,Y^\dagger\, Y\;.
\end{equation}
Here the $Y^\dagger\, Y$ term represents operators built up from  Yukawa
matrices transforming appropriately under the classical flavor symmetry
$G_\mathrm{flavor}~=~ \SU3_u \times \SU3_d \times \SU3_q \times \SU3_e \times
\SU3_\ell$ that appears in the SM when all Yukawas are set to zero.  It turns
out that, if one imposes \eqref{eq:MFVansatz} at the GUT scale, the form of
$\widetilde{m}^2$ stays preserved under the renormalization group. Even more,
the coefficients $a$ and $b$ in \eqref{eq:MFVansatz} get driven to non-trivial
quasi-fixed points~\cite{Paradisi:2008qh,Colangelo:2008qp}{}, which makes it
practically impossible to distinguish experimentally between zero and non-zero
$b$, i.e.\ an mSUGRA ansatz or its MFV-inspired generalization, at the GUT
scale. Moreover, the supersymmetric CP phases get driven to
zero~\cite{Colangelo:2008qp}, thus relaxing the supersymmetric CP problems.
Hence, the $D_4$ flavor symmetry ensures phenomenological properties very
close to those of the so-called mSUGRA ansatz, which is known to evade the
phenomenological constraints. In summary, the $D_4$ symmetry
seems to represent an appropriate means to ameliorate or even avoid the
supersymmetric flavor problems, without the need to rely on a specific scenario
of mediation of supersymmetry breaking. 

\subsection{Scale of supersymmetry breakdown and moduli stabilization}

This brings us to another very important question: how is supersymmetry broken
and why is the weak scale so far below the scale where gauge couplings meet? The
traditional answer to these questions relies on dimensional
transmutation~\cite{Witten:1981nf}{}, i.e.\ supersymmetry is broken by some
hidden sector that gets strong at an intermediate scale $\Lambda$. The gravitino
mass, setting the scale for the MSSM soft masses, is then given by
\cite{Nilles:1982ik}{}
\begin{equation}\label{eq:Lambda}
 m_{3/2}~\sim~\frac{\Lambda^3}{M_\mathrm{P}^2}\;.
\end{equation} 
However, if one is to embed this attractive scheme into string theory, one first
has to fix the moduli, in particular the dilaton, whose VEV sets the gauge
coupling and thus determines the scale of hidden sector strong dynamics
$\Lambda$. Often this re-introduces the problem of
hierarchies.\footnote{Exceptions to this statement are the race-track
scheme~\cite{Krasnikov:1987jj}{}, where one needs two hidden sectors with
rather special properties, and the K\"ahler stabilization
mechanism~\cite{Binetruy:1996xj,Casas:1996zi}{}
(for a review see \cite{Gaillard:2007jr}), which requires very favorable values
of certain parameters.} This is perhaps most transparent in the effective
superpotential obtained in the framework of flux compactifications
(a.k.a.\ KKLT \cite{Kachru:2003aw} superpotential)
\begin{equation}\label{eq:WKKLT}
 \mathscr{W}_\mathrm{KKLT}~=~c+A\,\mathrm{e}^{-a\,S}\;,
\end{equation}
where $c$ is a constant and the second term represents the hidden sector strong
dynamics, and $\re S \propto 1/g^2$. In the minimum, the second term adjusts its
size to $c$. In particular, the scale of the gravitino mass is set by $c$,
\begin{equation}
 m_{3/2}~\sim~c
\end{equation}
in Planck units. In the landscape
picture~\cite{Quevedo:2003cs,Susskind:2003kw}{} $c$ happens to be small by
anthropic reasons, i.e.\ although the natural scale for $c$ in flux
compactifications is order one in Planck units, due to a huge number of vacua
there are some with strongly suppressed $c$, and we happen to live in such a
vacuum.

In~\cite{Kappl:2008ie}{} an alternative has been proposed where $c$ emerges as the
VEV of the perturbative superpotential and its smallness is explained by a
symmetry (and hence in agreement with the more traditional criteria of
``naturalness''~\cite{'tHooft:1979bh}{}). It turns out that $R$ symmetries allow
us to control the VEV of the superpotential. First, a continuous $\U1_R$ implies
that, for configurations that satisfy the global supersymmetry $F$ term
equations of a polynomial, perturbative superpotential $\mathscr{W}_\mathrm{pert}$
\begin{equation}\label{eq:globalF}
 F_i~:=~\frac{\partial\mathscr{W}_\mathrm{pert}}{\partial\phi_i}~=~0\;,
\end{equation}
the expectation value of the superpotential vanishes~\cite{Kappl:2008ie}{},
\begin{equation}
 \langle\mathscr{W}_\mathrm{pert}\rangle~=~0\;.
\end{equation}
This statement holds regardless of whether $\U1_R$ is unbroken, where the
statement is trivial, or (spontaneously) broken.
Further, in the presence of an approximate $\U1_R$ symmetry, this statement gets
modified to 
\begin{equation}
 \langle\mathscr{W}_\mathrm{pert}\rangle~\sim~\langle\phi\rangle^N\;,
\end{equation}
where $N$ is the order at which explicit $R$ symmetry breaking terms appear and
$\langle\phi\rangle$ denotes the typical size of field VEVs. As it turns out,
orbifold models give us approximate $\U1_R$ symmetries. They are a consequence of
exact, discrete $R$ symmetries, reflecting a discrete rotational symmetry of
compact 6D space. Specifically, in the \Z6-II orbifold based on the Lie lattice
$\Lambda_{\G2\times\SU3\times\SO4}$, one has a
\begin{equation}\label{eq:DiscreteR}
 [\Z6\times\Z3\times\Z2]_R
\end{equation}
symmetry~\cite{Kobayashi:2004ya,Araki:2007ss}{}; other orbifolds have similar
discrete symmetries. Some vacua of the mini-landscape models have been analyzed;
the result is that the expectation value of the perturbative superpotential is
\begin{equation}
\langle\mathscr{W}_\mathrm{pert}\rangle~=~10^{-\mathcal{O}(10)}\;.
\end{equation}
Since the $R$ symmetry is only approximate, the notoriously troublesome $R$ axion is
massive and therefore harmless. One retains instead an approximate $R$ axion
$\eta$, whose mass is slightly enhanced against against $\langle\mathscr{W}_\mathrm{pert}\rangle$,
\begin{equation}
 m_\eta~\simeq~\frac{\langle\mathscr{W}_\mathrm{pert}\rangle}{\langle\phi\rangle^2}\;. 
\end{equation}
Further, in many configurations, this is the only light mode, i.e.\ the curvature
in all other directions is much larger. Explicit examples for simple
field-theoretic toy models as well as a description of the situation for the
mini-landscape models have been discussed very recently~\cite{Brummer:2010fr}{}.

Let us comment that approximate continuous symmetries, which arise from
high-power exact discrete symmetries, might also allow us to solve other
fundamental problems, such as the strong CP problem~\cite{Choi:2009jt}{}.

The hierarchically small vacuum expectation value of the superpotential can be
used in order to stabilize the dilaton~\cite{Kappl:2008ie}{}. One obtains an
effective, KKLT-like superpotential,
\begin{equation}
 \mathscr{W}_\mathrm{eff}
 ~=~
 \langle\mathscr{W}_\mathrm{pert}\rangle+A\,\mathrm{e}^{-a\,S}+\frac{1}{2}m_\eta\,\eta^2\;,
\end{equation}
which describes the `hidden sector' up to heavier modes. In the supersymmetric
minimum, the non-perturbative term $A\,\mathrm{e}^{-a\,S}$ adjusts its size to
the VEV of the perturbative superpotential $\langle\mathscr{W}_\mathrm{pert}\rangle$. At this
level, one obtains a vacuum with a fixed dilaton.\footnote{The vacuum energy
depends on the K\"ahler potential of the matter
fields~\cite[section~4]{Weinberg:1988cp}{}.} The vacuum
expectation value of the superpotential, i.e.\ the gravitino mass, is given by
\begin{equation}
 m_{3/2}~\simeq~\langle\mathscr{W}_\mathrm{eff}\rangle~\sim~\langle\mathscr{W}_\mathrm{pert}\rangle
 \;.
\end{equation}
As usual, $m_{3/2}$ will set the scale for the MSSM soft masses and the
electroweak scale.
One may also speculate that some hidden sector matter fields acquire $F$-terms
such as to to correct the vacuum energy in the spirit of `$F$-term
uplifting'~\cite{Lebedev:2006qq}{}; a possible way to stabilize the $T$-moduli
has been sketched in \cite{Dundee:2010sb}. This might then lead to a `mirage-like'
pattern of soft supersymmetric masses~\cite{Lowen:2008fm}{}, and will be
discussed in more detail elsewhere.

An important question concerns whether the size of the non-perturbative term is
consistent with a phenomenologically viable gauge coupling $g$, i.e.\ whether
the dilaton gets fixed at realistic values. This question can be answered
affirmatively. In order to see this, let us briefly review the analysis of
\cite{Lebedev:2006tr}, where the scale $\Lambda$ of hidden sector strong
dynamics in the mini-landscape has been studied.  Here, the hidden sector is
defined as the gauge group with the largest $\beta$-function coefficient. It was
found that, for a realistic gauge coupling, $\Lambda$ takes values which are, by
the relation \eqref{eq:Lambda}, consistent with TeV-scale soft masses. 
\begin{figure}
\centerline{
\subfigure[2~WL (from~\cite{Lebedev:2006tr}{}).]{\CenterObject{\includegraphics{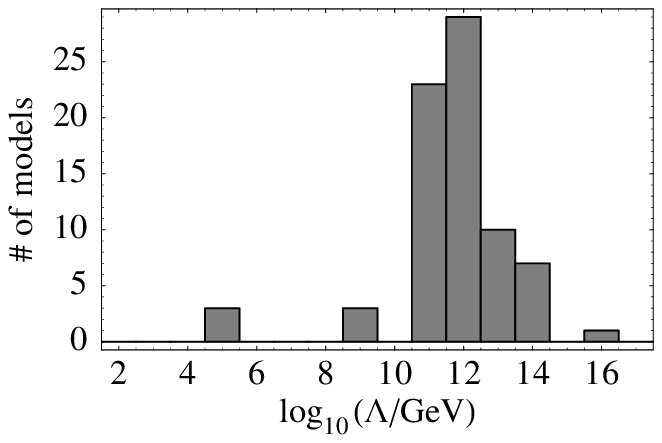}}}
\quad
\subfigure[3~WL (from~\cite{Lebedev:2008un}{}).]{\CenterObject{\includegraphics[scale=0.7]{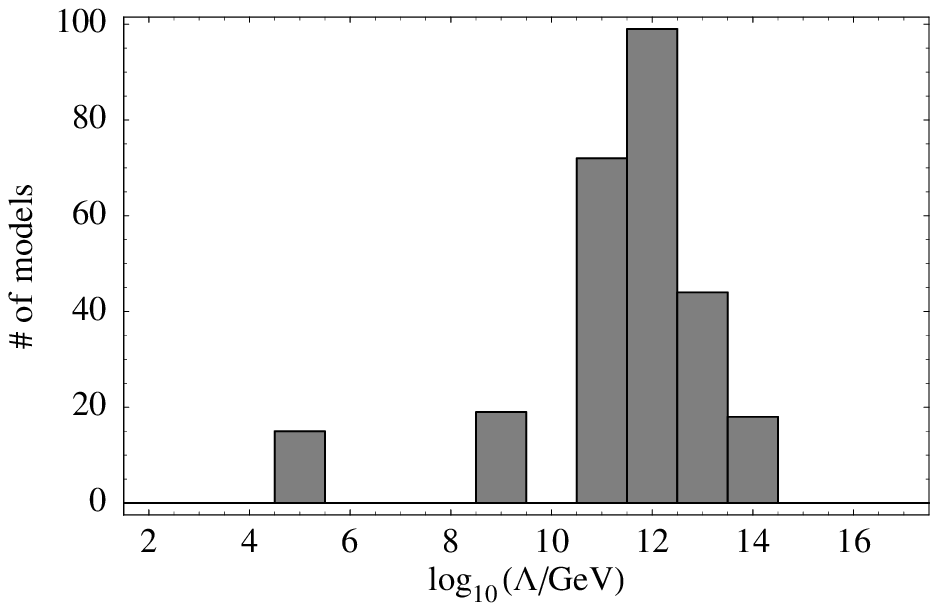}}}
}
\caption{Statistics of the scale of hidden sector gaugino condensation $\Lambda$
in (a) the heterotic mini-landscape and (b) its extension to three Wilson
lines.  $\Lambda$ is peaked at values, where according to \eqref{eq:Lambda}
the soft masses are in the TeV range.}
\label{fig:statistics}
\end{figure}
This is illustrated in figure~\ref{fig:statistics}, where we also show the
result of the completion of the mini-landscape search~\cite{Lebedev:2008un}{}.
These statistics show that, under the assumption of a realistic gauge coupling,
$\Lambda$ is such that by relation~\eqref{eq:Lambda} a phenomenologically
attractive gravitino mass emerges. Turning this argument around, we see that
once we are in a vacuum with $\langle\mathscr{W}_\mathrm{pert}\rangle\sim10^{-15}$, the hidden
sector $\beta$-function coefficients are such that the dilaton gets fixed at
a realistic value.

The hierarchically small vacuum expectation value of $\mathscr{W}_\mathrm{pert}$ has important
consequences for the solution of the MSSM $\mu$ problem. In these models, 
$\langle\mathscr{W}_\mathrm{pert}\rangle$ is proportional to the MSSM $\mu$
parameter~\cite{Casas:1992mk}{}. This can be shown by expanding the K\"ahler
potential~\cite{Antoniadis:1994hg}{} in the holomorphic combination
$h_u\,h_d$~\cite{Brummer:2010fr}. One then obtains a which correspond to
holomorphic~\cite{Kim:1983dt}{} term proportional to the VEV of the
superpotential and Giudice-Masiero type~\cite{Giudice:1988yz}{}
contribution, so altogether 
\begin{equation}
 \mu~\sim~m_{3/2}\;.
\end{equation}
Taking into account both contributions, this can lead to consistent boundary
conditions for the soft masses. This has been discussed
recently in the framework of 5D orbifold GUTs~\cite{Brummer:2009ug}, where the
possible contribution of a Chern-Simons term to the $\mu$ term has been taken
into account. It turns out that the Chern-Simons contribution is crucial in
order to obtain a viable phenomenology~\cite{Brummer:2009ug}{}. It appears also
interesting to see if, by having a better understanding of the origin of the
MSSM $\mu$ parameter, one might be able to shed some light on the MSSM fine
tuning problem.
 
Before summarizing the good features in the last section, let us briefly comment
on open questions and potential problems. There might be a mild tension between
the estimated size of the coefficients of dimension five proton decay operators
and the observed proton longevity; on the other hand, we have not really
obtained a full understanding of the patterns of the Yukawa couplings. It might
well turn out that, once we fully understand why the $u$ and $d$ Yukawa
couplings are so small, we will also be able to explain why the first
generations coefficients of the $qqq\ell$ and $\bar u\bar u\bar d \bar e$
operators are highly suppressed (cf.\ also~\cite{Smith:2008ju} for a recent,
very similar discussion).\footnote{The author is grateful to E.~Witten for
stressing this.} We have further argued that highly anisotropic
compactifications might allow us to reconcile the discrepancy between the string
and GUT scales. However, so far we do not have obtained a dynamical mechanism
that allows us to understand why there is a hierarchy between the radii. And, of
course, we have not much to say on the most fundamental questions such as the
observed vacuum energy.

\section{Summary} 

In these proceedings, some progress of embedding ideas of grand unification into
string theory is described. Field-theoretic orbifold GUTs provided us with the
geometric intuition for how to efficiently search for realistic models. This has
lead to the concept of local grand unification which gives a simple explanation
for the simultaneous existence of complete GUT multiplets and split multiplets
in Nature. Using this as a guideline, potentially realistic models with the
exact MSSM spectra and a simple geometric interpretation have been obtained.
These models have vacua with $R$ parity and are consistent with MSSM gauge
coupling unification. That is, we imposed our prejudices of supersymmetry and
unification in our model search, where we had to disregard models which are not
consistent with our criteria. Amazingly, those of the models which survive this
selection process have many unexpected features, which were not `put in by hand'
but happen to occur automatically. The most striking of these `stringy
surprises' are:
\begin{dinglist}{"50}
 \item neutrino see-saw with the see-saw scale somewhat below the GUT or
 compactification scale;
 \item ``gauge-top unification'' and a correlation between reasonable values
 for $\tan\beta$ and anisotropy of the model;
 \item a potential solution to the supersymmetric flavor and CP problems based on the
 non-Abelian discrete flavor symmetry $D_4$;
 \item high-power discrete $R$ symmetries explaining a hierarchically small
 gravitino mass;
 \item a hidden sector whose scale of strong dynamics is consistent with a TeV
 scale gravitino mass;
 \item a relation between the $\mu$ term and the scale of supersymmetry
 breakdown.
\end{dinglist}
Future might tell us whether these are just accidents or connected to the real
world.

\subsection*{Acknowledgments}

It is a pleasure to thank
 F.~Br\"ummer,
 W.~Buchm\"uller, K.~Hamaguchi, 
 P.~Hosteins,
 R.~Kappl, T.~Kobayashi, 
 O.~Lebedev, H.P.~Nilles, 
 P.~Paradisi, F.~Pl\"oger, 
 S.~Raby, S.~Ramos-S\'anchez, 
 R.~Schieren, K.~Schmidt-Hoberg, 
 C.~Simonetto and
 P.~Vaudrevange 
for very fruitful collaborations, and K.~Schmidt-Hoberg and P.~Vaudrevange for
comments.
Further thanks go to the organizers of the Yukawa symposium for the wonderful
meeting and the fantastic conference dinner.
This research was supported by the DFG cluster of excellence Origin and
Structure of the Universe and the \mbox{SFB-Transregio} 27 "Neutrinos and
Beyond" by Deutsche Forschungsgemeinschaft (DFG).

\bibliography{Orbifold}

\providecommand{\bysame}{\leavevmode\hbox to3em{\hrulefill}\thinspace}
\begin{thebibliography}{100}

\bibitem{Olive:1981tb}
D.~I. Olive, Invited talk given at Study Conf. on Unification of Fundamental
  Interactions II, Erice, Italy, Oct 6-14, 1981.

\bibitem{Gross:1984dd}
D.~J. Gross, J.~A. Harvey, E.~J. Martinec, and R.~Rohm, Phys. Rev. Lett.
  \textbf{54} (1985), 502--505.

\bibitem{Gross:1985fr}
D.~J. Gross, J.~A. Harvey, E.~J. Martinec, and R.~Rohm, Nucl. Phys.
  \textbf{B256} (1985), 253.

\bibitem{Nilles:2004ej}
H.~P. Nilles,  (2004),  hep-th/0410160.

\bibitem{Ibanez:1987dw}
L.~E. Ib{\'a}{\~n}ez,  (1987), Based on lectures given at the XVII GIFT Seminar
  on Strings and Superstrings, El Escorial, Spain, Jun 1-6, 1987 and Mt. Sorak
  Symposium, Korea, Jul 1987 and ELAF '87, La Plata, Argentina, Jul 6-24, 1987.

\bibitem{Ratz:2007my}
M.~Ratz,  (2007),  arXiv:0711.1582 [hep-ph].

\bibitem{Nilles:2008gq}
H.~P. Nilles, S.~Ramos-S{\'a}nchez, M.~Ratz, and P.~K.~S. Vaudrevange, Eur.
  Phys. J. \textbf{C59} (2009), 249--267,  [0806.3905].

\bibitem{Raby:2008gh}
S.~Raby, Eur. Phys. J. \textbf{C59} (2009), 223--247,  [0807.4921].

\bibitem{Gmeiner:2007zz}
F.~Gmeiner and G.~Honecker, JHEP \textbf{09} (2007), 128,  [arXiv:0708.2285
  [hep-th]].

\bibitem{Gmeiner:2008xq}
F.~Gmeiner and G.~Honecker, JHEP \textbf{07} (2008), 052,  [0806.3039].

\bibitem{Cleaver:2007ek}
G.~B. Cleaver,  (2007),  hep-ph/0703027.

\bibitem{Donagi:2008xy}
R.~Donagi and K.~Wendland,  (2008),  0809.0330.

\bibitem{Bouchard:2005ag}
V.~Bouchard and R.~Donagi, Phys. Lett. \textbf{B633} (2006), 783--791,
  [hep-th/0512149].

\bibitem{Braun:2005nv}
V.~Braun, Y.-H. He, B.~A. Ovrut, and T.~Pantev, JHEP \textbf{05} (2006), 043,
  [hep-th/0512177].

\bibitem{Kim:2007mt}
J.~E. Kim, J.-H. Kim, and B.~Kyae, JHEP \textbf{06} (2007), 034,
  [hep-ph/0702278].

\bibitem{Buchmuller:2005jr}
W.~Buchm{\"u}ller, K.~Hamaguchi, O.~Lebedev, and M.~Ratz, Phys. Rev. Lett.
  \textbf{96} (2006), 121602,  [hep-ph/0511035].

\bibitem{Buchmuller:2006ik}
W.~Buchm{\"u}ller, K.~Hamaguchi, O.~Lebedev, and M.~Ratz, Nucl. Phys.
  \textbf{B785} (2007), 149--209,  [hep-th/0606187].

\bibitem{Lebedev:2006kn}
O.~Lebedev, H.~P. Nilles, S.~Raby, S.~Ramos-S{\'a}nchez, M.~Ratz, P.~K.~S.
  Vaudrevange, and A.~Wingerter, Phys. Lett. \textbf{B645} (2007), 88,
  [hep-th/0611095].

\bibitem{Lebedev:2006tr}
O.~Lebedev, H.~P. Nilles, S.~Raby, S.~Ramos-S{\'a}nchez, M.~Ratz, P.~K.~S.
  Vaudrevange, and A.~Wingerter, Phys. Rev. Lett. \textbf{98} (2007), 181602,
  [hep-th/0611203].

\bibitem{Lebedev:2007hv}
O.~Lebedev, H.~P. Nilles, S.~Raby, S.~Ramos-S{\'a}nchez, M.~Ratz, P.~K.~S.
  Vaudrevange, and A.~Wingerter, Phys. Rev. \textbf{D77} (2007), 046013,
  [arXiv:0708.2691 [hep-th]].

\bibitem{Lebedev:2008un}
O.~Lebedev, H.~P. Nilles, S.~Ramos-S\'{a}nchez, M.~Ratz, and P.~K.~S.
  Vaudrevange, Phys. Lett. \textbf{B668} (2008), 331--335,  [0807.4384].

\bibitem{Kawamura:1999nj}
Y.~Kawamura, Prog. Theor. Phys. \textbf{103} (2000), 613--619,
  [hep-ph/9902423].

\bibitem{Kawamura:2000ev}
Y.~Kawamura, Prog. Theor. Phys. \textbf{105} (2001), 999--1006,
  [hep-ph/0012125].

\bibitem{Altarelli:2001qj}
G.~Altarelli and F.~Feruglio, Phys. Lett. \textbf{B511} (2001), 257--264,
  [hep-ph/0102301].

\bibitem{Hall:2001pg}
L.~J. Hall and Y.~Nomura, Phys. Rev. \textbf{D64} (2001), 055003,
  [hep-ph/0103125].

\bibitem{Hebecker:2001wq}
A.~Hebecker and J.~March-Russell, Nucl. Phys. \textbf{B613} (2001), 3--16,
  [hep-ph/0106166].

\bibitem{Asaka:2001eh}
T.~Asaka, W.~Buchm{\"u}ller, and L.~Covi, Phys. Lett. \textbf{B523} (2001),
  199--204,  [hep-ph/0108021].

\bibitem{Hall:2001xr}
L.~J. Hall, Y.~Nomura, T.~Okui, and D.~R. Smith, Phys. Rev. \textbf{D65}
  (2002), 035008,  [hep-ph/0108071].

\bibitem{Burdman:2002se}
G.~Burdman and Y.~Nomura, Nucl. Phys. \textbf{B656} (2003), 3--22,
  [hep-ph/0210257].

\bibitem{Quiros:2003gg}
M.~Quiros,  (2003),  hep-ph/0302189.

\bibitem{Dixon:1985jw}
L.~J. Dixon, J.~A. Harvey, C.~Vafa, and E.~Witten, Nucl. Phys. \textbf{B261}
  (1985), 678--686.

\bibitem{Dixon:1986jc}
L.~J. Dixon, J.~A. Harvey, C.~Vafa, and E.~Witten, Nucl. Phys. \textbf{B274}
  (1986), 285--314.

\bibitem{Ibanez:1986tp}
L.~E. Ib{\'a}{\~n}ez, H.~P. Nilles, and F.~Quevedo, Phys. Lett. \textbf{B187}
  (1987), 25--32.

\bibitem{Ibanez:1987sn}
L.~E. Ib{\'a}{\~n}ez, J.~E. Kim, H.~P. Nilles, and F.~Quevedo, Phys. Lett.
  \textbf{B191} (1987), 282--286.

\bibitem{Casas:1987us}
J.~A. Casas, E.~K. Katehou, and C.~Mu{\~{n}}oz, Nucl. Phys. \textbf{B317}
  (1989), 171.

\bibitem{Casas:1988hb}
J.~A. Casas and C.~Mu{\~n}oz, Phys. Lett. \textbf{B214} (1988), 63.

\bibitem{Font:1988tp}
A.~Font, L.~E. Ib{\'a}{\~n}ez, H.~P. Nilles, and F.~Quevedo, Nucl. Phys.
  \textbf{B307} (1988), 109, Erratum {\em ibid.} {\bf B310}.

\bibitem{Font:1988nc}
A.~Font, L.~E. Ib{\'a}{\~n}ez, H.~P. Nilles, and F.~Quevedo, Phys. Lett.
  \textbf{B213} (1988), 274.

\bibitem{Forste:2004ie}
S.~F{\"o}rste, H.~P. Nilles, P.~K.~S. Vaudrevange, and A.~Wingerter, Phys. Rev.
  \textbf{D70} (2004), 106008,  [hep-th/0406208].

\bibitem{Buchmuller:2005sh}
W.~Buchm{\"u}ller, K.~Hamaguchi, O.~Lebedev, and M.~Ratz,  (2005),
  hep-ph/0512326.

\bibitem{Buchmuller:2007qf}
W.~Buchm{\"u}ller, C.~L{\"u}deling, and J.~Schmidt, JHEP \textbf{09} (2007),
  113,  [arXiv:0707.1651 [hep-ph]].

\bibitem{Pati:1974yy}
J.~C. Pati and A.~Salam, Phys. Rev. \textbf{D10} (1974), 275--289.

\bibitem{Schmidt:2009bd}
J.~Schmidt,  (2009),  0906.5501.

\bibitem{Vaudrevange:2008sm}
P.~K.~S. Vaudrevange,  (2008),  0812.3503.

\bibitem{RamosSanchez:2008tn}
S.~Ramos-S{\'a}nchez,  (2008),  0812.3560.

\bibitem{Giedt:2000bi}
J.~Giedt, Ann. Phys. \textbf{289} (2001), 251,  [hep-th/0009104].

\bibitem{Giedt:2003an}
J.~Giedt, Nucl. Phys. \textbf{B671} (2003), 133--147,  [hep-th/0301232].

\bibitem{Ploger:2007iq}
F.~Pl{\"o}ger, S.~Ramos-S{\'a}nchez, M.~Ratz, and P.~K.~S. Vaudrevange, JHEP
  \textbf{04} (2007), 063,  [hep-th/0702176].

\bibitem{Dienes:2006ut}
K.~R. Dienes, Phys. Rev. \textbf{D73} (2006), 106010,  [hep-th/0602286].

\bibitem{Dienes:2006ca}
K.~R. Dienes and M.~Lennek, Phys. Rev. \textbf{D75} (2007), 026008,
  [hep-th/0610319].

\bibitem{Buchmuller:2004hv}
W.~Buchm{\"u}ller, K.~Hamaguchi, O.~Lebedev, and M.~Ratz, Nucl. Phys.
  \textbf{B712} (2005), 139--156,  [hep-ph/0412318].

\bibitem{Kobayashi:2004ud}
T.~Kobayashi, S.~Raby, and R.-J. Zhang, Phys. Lett. \textbf{B593} (2004),
  262--270,  [hep-ph/0403065].

\bibitem{Kobayashi:2004ya}
T.~Kobayashi, S.~Raby, and R.-J. Zhang, Nucl. Phys. \textbf{B704} (2005),
  3--55,  [hep-ph/0409098].

\bibitem{Buchmuller:2008uq}
W.~Buchm{\"u}ller and J.~Schmidt, Nucl. Phys. \textbf{B807} (2009), 265--289,
  [0807.1046].

\bibitem{GrootNibbelink:2007pn}
S.~{Groot Nibbelink}, T.-W. Ha, and M.~Trapletti, Phys. Rev. \textbf{D77}
  (2008), 026002,  [arXiv:0707.1597 [hep-th]].

\bibitem{Nibbelink:2008tv}
S.~{Groot Nibbelink}, D.~Klevers, F.~Pl{\"o}ger, M.~Trapletti, and P.~K.~S.
  Vaudrevange, JHEP \textbf{04} (2008), 060,  [0802.2809].

\bibitem{Nibbelink:2009sp}
S.~{Groot Nibbelink}, J.~Held, F.~Ruehle, M.~Trapletti, and P.~K.~S.
  Vaudrevange, JHEP \textbf{03} (2009), 005,  [0901.3059].

\bibitem{Dine:1998up}
M.~Dine, Prog. Theor. Phys. Suppl. \textbf{134} (1999), 1--17,
  [hep-th/9903212].

\bibitem{Kofman:2004yc}
L.~Kofman et~al., JHEP \textbf{05} (2004), 030,  [hep-th/0403001].

\bibitem{Buchmuller:2009er}
W.~Buchm{\"u}ller, R.~Catena, and K.~Schmidt-Hoberg,  (2009),  0902.4512.

\bibitem{Atick:1987gy}
J.~J. Atick, L.~J. Dixon, and A.~Sen, Nucl. Phys. \textbf{B292} (1987),
  109--149.

\bibitem{Minkowski:1977sc}
P.~Minkowski, Phys. Lett. \textbf{B67} (1977), 421.

\bibitem{Buchmuller:2007zd}
W.~Buchm{\"u}ller, K.~Hamaguchi, O.~Lebedev, S.~Ramos-S{\'a}nchez, and M.~Ratz,
  Phys. Rev. Lett. \textbf{99} (2007), 021601,  [hep-ph/0703078].

\bibitem{Ellis:2007wz}
J.~R. Ellis and O.~Lebedev, Phys. Lett. \textbf{B653} (2007), 411--418,
  [arXiv:0707.3419 [hep-ph]].

\bibitem{Hall:1999sn}
L.~J. Hall, H.~Murayama, and N.~Weiner, Phys. Rev. Lett. \textbf{84} (2000),
  2572--2575,  [hep-ph/9911341].

\bibitem{Smirnov:1993af}
A.~Y. Smirnov, Phys. Rev. \textbf{D48} (1993), 3264--3270,  [hep-ph/9304205].

\bibitem{Lee:2003mc}
H.~M. Lee, H.~P. Nilles, and M.~Zucker, Nucl. Phys. \textbf{B680} (2004),
  177--198,  [hep-th/0309195].

\bibitem{Hosteins:2009xk}
P.~Hosteins, R.~Kappl, M.~Ratz, and K.~Schmidt-Hoberg, JHEP \textbf{07} (2009),
  029,  [0905.3323].

\bibitem{Witten:1996mz}
E.~Witten, Nucl. Phys. \textbf{B471} (1996), 135--158,  [hep-th/9602070].

\bibitem{Hebecker:2004ce}
A.~Hebecker and M.~Trapletti, Nucl. Phys. \textbf{B713} (2005), 173--203,
  [hep-th/0411131].

\bibitem{Dundee:2008ts}
B.~Dundee, S.~Raby, and A.~Wingerter,  (2008),  0805.4186.

\bibitem{Dundee:2008gr}
B.~Dundee, S.~Raby, and A.~Wingerter,  (2008),  0811.4026.

\bibitem{Kim:2007jg}
J.~E. Kim and B.~Kyae,  (2007),  0712.1596.

\bibitem{Klaput:2010dg}
M.~A. Klaput and C.~Paleani,  (2010),  1001.1480.

\bibitem{Kobayashi:2006wq}
T.~Kobayashi, H.~P. Nilles, F.~Pl{\"o}ger, S.~Raby, and M.~Ratz, Nucl. Phys.
  \textbf{B768} (2007), 135--156,  [hep-ph/0611020].

\bibitem{Ko:2007dz}
P.~Ko, T.~Kobayashi, J.-h. Park, and S.~Raby, Phys. Rev. \textbf{D76} (2007),
  035005,  [arXiv:0704.2807 [hep-ph]].

\bibitem{Chivukula:1987py}
R.~S. Chivukula and H.~Georgi, Phys. Lett. \textbf{B188} (1987), 99.

\bibitem{Buras:2000dm}
A.~J. Buras, P.~Gambino, M.~Gorbahn, S.~J{\"a}ger, and L.~Silvestrini, Phys.
  Lett. \textbf{B500} (2001), 161--167,  [hep-ph/0007085].

\bibitem{D'Ambrosio:2002ex}
G.~D'Ambrosio, G.~F. Giudice, G.~Isidori, and A.~Strumia, Nucl. Phys.
  \textbf{B645} (2002), 155--187,  [hep-ph/0207036].

\bibitem{Paradisi:2008qh}
P.~Paradisi, M.~Ratz, R.~Schieren, and C.~Simonetto, Phys. Lett. \textbf{B668}
  (2008), 202--209,  [0805.3989].

\bibitem{Colangelo:2008qp}
G.~Colangelo, E.~Nikolidakis, and C.~Smith, Eur. Phys. J. \textbf{C59} (2009),
  75--98,  [0807.0801].

\bibitem{Witten:1981nf}
E.~Witten, Nucl. Phys. \textbf{B188} (1981), 513.

\bibitem{Nilles:1982ik}
H.~P. Nilles, Phys. Lett. \textbf{B115} (1982), 193.

\bibitem{Krasnikov:1987jj}
N.~V. Krasnikov, Phys. Lett. \textbf{B193} (1987), 37--40.

\bibitem{Binetruy:1996xj}
P.~Bin{\'e}truy, M.~K. Gaillard, and Y.-Y. Wu, Nucl. Phys. \textbf{B481}
  (1996), 109--128,  [hep-th/9605170].

\bibitem{Casas:1996zi}
J.~A. Casas, Phys. Lett. \textbf{B384} (1996), 103--110,  [hep-th/9605180].

\bibitem{Gaillard:2007jr}
M.~K. Gaillard and B.~D. Nelson, Int. J. Mod. Phys. \textbf{A22} (2007), 1451,
  [hep-th/0703227].

\bibitem{Kachru:2003aw}
S.~Kachru, R.~Kallosh, A.~Linde, and S.~P. Trivedi, Phys. Rev. \textbf{D68}
  (2003), 046005,  [hep-th/0301240].

\bibitem{Quevedo:2003cs}
F.~Quevedo, Phys. World \textbf{16N11} (2003), 21--22.

\bibitem{Susskind:2003kw}
L.~Susskind,  (2003),  hep-th/0302219.

\bibitem{Kappl:2008ie}
R.~Kappl, H.~P. Nilles, S.~Ramos-S{\'a}nchez, M.~Ratz, K.~Schmidt-Hoberg, and
  P.~K. Vaudrevange, Phys. Rev. Lett. \textbf{102} (2009), 121602,
  [0812.2120].

\bibitem{'tHooft:1979bh}
G.~'t~Hooft, NATO Adv. Study Inst. Ser. B Phys. \textbf{59} (1980), 135.

\bibitem{Araki:2007ss}
T.~Araki, K.-S. Choi, T.~Kobayashi, J.~Kubo, and H.~Ohki,  (2007),
  arXiv:0705.3075 [hep-ph].

\bibitem{Brummer:2010fr}
F.~Br{\"u}mmer, R.~Kappl, M.~Ratz, and K.~Schmidt-Hoberg,  (2010),  1003.0084.

\bibitem{Choi:2009jt}
K.-S. Choi, H.~P. Nilles, S.~Ramos-S{\'a}nchez, and P.~K.~S. Vaudrevange,
  (2009),  0902.3070.

\bibitem{Weinberg:1988cp}
S.~Weinberg, Rev. Mod. Phys. \textbf{61} (1989), 1--23.

\bibitem{Lebedev:2006qq}
O.~Lebedev, H.~P. Nilles, and M.~Ratz, Phys. Lett. \textbf{B636} (2006),
  126--131,  [hep-th/0603047].

\bibitem{Dundee:2010sb}
B.~Dundee, S.~Raby, and A.~Westphal,  (2010),  1002.1081.

\bibitem{Lowen:2008fm}
V.~L{\"o}wen and H.~P. Nilles, Phys. Rev. \textbf{D77} (2008), 106007,
  [0802.1137].

\bibitem{Casas:1992mk}
J.~A. Casas and C.~Mu{\~n}oz, Phys. Lett. \textbf{B306} (1993), 288--294,
  [hep-ph/9302227].

\bibitem{Antoniadis:1994hg}
I.~Antoniadis, E.~Gava, K.~S. Narain, and T.~R. Taylor, Nucl. Phys.
  \textbf{B432} (1994), 187--204,  [hep-th/9405024].

\bibitem{Kim:1983dt}
J.~E. Kim and H.~P. Nilles, Phys. Lett. \textbf{B138} (1984), 150.

\bibitem{Giudice:1988yz}
G.~F. Giudice and A.~Masiero, Phys. Lett. \textbf{B206} (1988), 480--484.

\bibitem{Brummer:2009ug}
F.~Br{\"u}mmer, S.~Fichet, A.~Hebecker, and S.~Kraml, JHEP \textbf{08} (2009),
  011,  [0906.2957].

\bibitem{Smith:2008ju}
C.~Smith,  (2008),  0809.3152.

\end{thebibliography}
\bibliographystyle{ArXiv}

\end{document}